\documentclass[aps,prb,reprint,showpacs,twocolumn,superscriptaddress,floatfix]{revtex4-1}

\usepackage[T1]{fontenc}
\usepackage[utf8]{inputenc}
\usepackage{epsfig}
\usepackage{amsmath}
\usepackage{amssymb}
\usepackage{amsfonts}
\usepackage{mathptmx}
\usepackage{eucal}
\usepackage{bm}
\usepackage{color}
\usepackage[colorlinks,linkcolor=blue,citecolor=blue]{hyperref}
\usepackage{epstopdf}
\usepackage{graphicx}
\begin{document}



\title{Negative differential thermal conductance and heat amplification in superconducting hybrid devices}

\author{Antonio Fornieri}
\email{antonio.fornieri@sns.it}
\affiliation{NEST, Istituto Nanoscienze-CNR and Scuola Normale Superiore, I-56127 Pisa, Italy}

\author{Giuliano Timossi}
\affiliation{NEST, Istituto Nanoscienze-CNR and Scuola Normale Superiore, I-56127 Pisa, Italy}

\author{Riccardo Bosisio}
\affiliation{SPIN-CNR, Via Dodecaneso 33, Genova I-16146, Italy}
\affiliation{NEST, Istituto Nanoscienze-CNR and Scuola Normale Superiore, I-56127 Pisa, Italy}

\author{Paolo Solinas}
\affiliation{SPIN-CNR, Via Dodecaneso 33, Genova I-16146, Italy}

\author{Francesco Giazotto}
\email{giazotto@sns.it}
\affiliation{NEST, Istituto Nanoscienze-CNR and Scuola Normale Superiore, I-56127 Pisa, Italy}




\begin{abstract}
We investigate the thermal transport properties of a temperature-biased Josephson tunnel junction composed of two different superconductors. We show that this simple system can provide a large negative differential thermal conductance (NDTC) with a peak-to-valley ratio of $\sim 3$ in the transmitted electronic heat current. The NDTC is then exploited to outline the caloritronic analogue of the tunnel diode, which can exhibit a modulation of the output temperature as large as 80 mK at a bath temperature of 50 mK. Moreover, this device may work in a regime of thermal hysteresis that can be used to store information as a thermal memory. On the other hand, the NDTC effect offers the opportunity to conceive two different designs of a thermal transistor, which might operate as a thermal switch or as an amplifier/modulator. The latter shows a heat amplification factor $>1$ in a 500-mK-wide working region of the gate temperature. After the successful realization of heat interferometers and thermal diodes, this kind of structures would complete the conversion of the most important electronic devices in their thermal counterparts, breaking ground for coherent caloritronics nanocircuits where heat currents can be manipulated at will.
\end{abstract}

\pacs{}

\maketitle


\section{Introduction}\label{sect1}
In the last decade an increasing interest has grown around the possibility to master thermal currents at the nanoscale with the same degree of accuracy obtained in contemporary electronic devices.\cite{GiazottoRev,Dubi,LiRev} This ability would benefit a great number of nanoscience fields, such as solid state cooling,\cite{GiazottoRev,QuarantaAPL,PekolaRev} thermal isolation,\cite{MartinezNatRect,FornieriAIP} radiation detection\cite{GiazottoRev} and quantum computing.\cite{NielsenChuang,Spilla} Although being still in their infancy, emerging fields like coherent caloritronics,\cite{GiazottoNature,MartinezRev} phononics and thermal logic \cite{LiRev} have already demonstrated remarkable results towards the implementation of the thermal counterparts of interferometers,\cite{GiazottoNature,MartinezNature,FornieriNature} diodes\cite{ChangScience,MartinezNatRect} and solid-state memory devices.\cite{XieAFM} Nevertheless, modern electronics had a phenomenal expansion only after the invention of the transistor,\cite{BardeenBrattain} whose thermal analogue remains one of the main goals to achieve the full control of heat currents and to finally realize thermal logic gates.\cite{LiRev}

Exactly ten years ago, Li and coworkers put forward the first theoretical proposal for a thermal transistor,\cite{LiWangCasatiAPL} indicating negative differential thermal conductance (NDTC) as an essential requirement to let the device work as a switch or an amplifier. Here, we show that a simple Josephson junction (JJ) between two different superconductors residing at different temperatures can provide a sizeable NDTC, which may give rise to various remarkable effects, like thermal hysteresis and heat amplification. As a result, we can envision several interesting non-linear devices to master electronic heat currents, including the thermal analogues of tunnel diodes,\cite{Esaki} memories\cite{XieAFM} and transistors.\cite{BardeenBrattain} The proposed devices could be realized with conventional nanofabrication techniques \cite{GiazottoNature,MartinezNature,MartinezNatRect,FornieriNature} and might be immediately exploited in low-temperature solid-state thermal circuits.

\begin{figure}[t]
\centering
\includegraphics[width=\columnwidth]{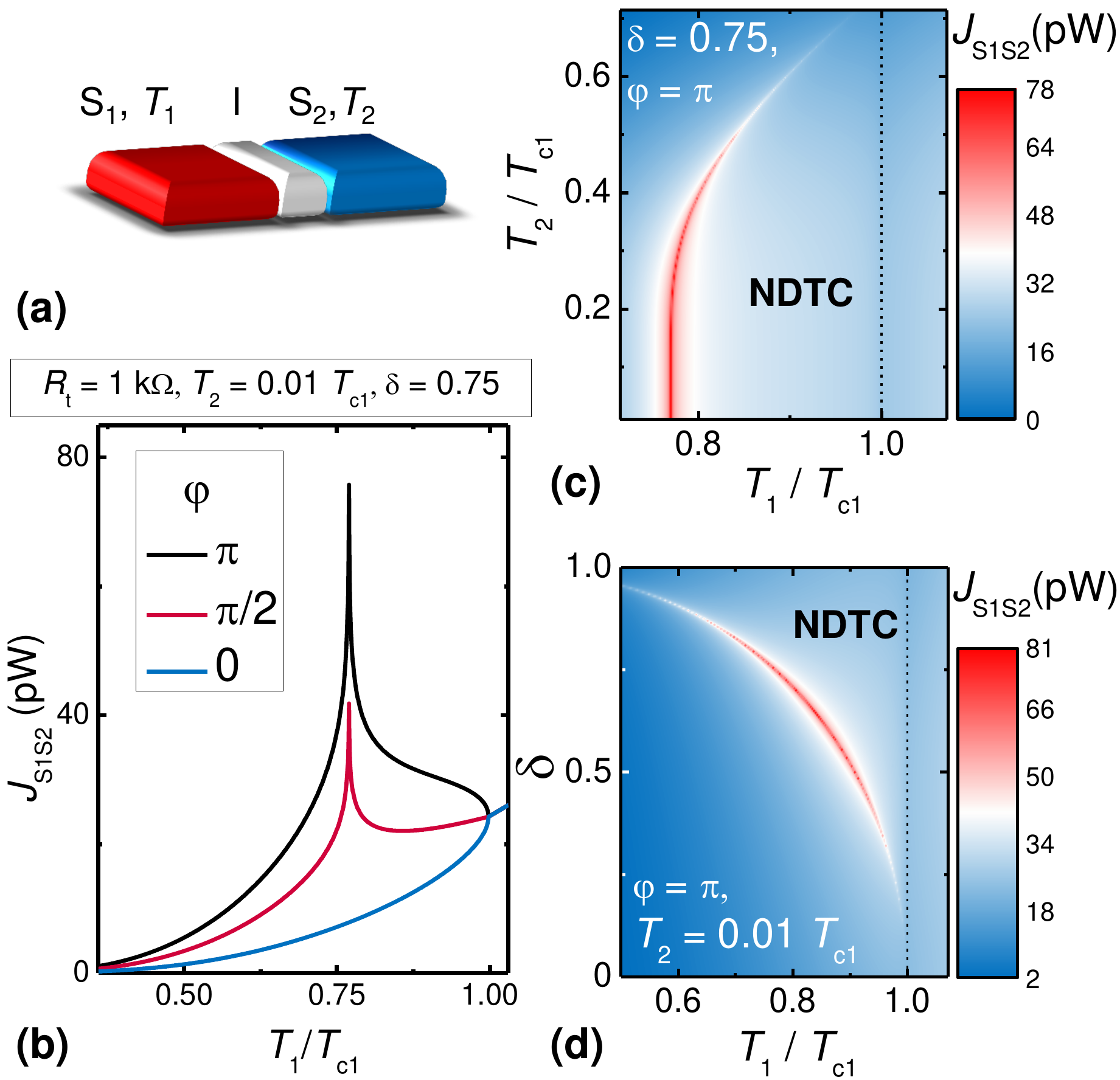}
\caption{Thermal transport through a JJ. (a) Schematic configuration of a JJ formed by two superconductors S$_1$ and S$_2$ at different temperatures $T_1$ and $T_2$, with $\delta=\Delta_2(0)/\Delta_1(0) \leq 1$. (b) Electronic heat current $J_{\rm S_1S_2}$ vs. $T_1$ at $T_2=0.01 T_{\rm c1}$ and for different values of the phase difference $\varphi$ between the superconducting condensates. 
All the curves are calculated for $\delta=0.75$. (c) Contour plot showing $J_{\rm S_1S_2}$ as a function of $T_1$ and $T_2$ for $\varphi=\pi$ and $\delta=0.75$. (d) Contour plot showing $J_{\rm S_1S_2}$ vs. $T_1$ and $\delta$ for $T_2=0.01T_{\rm c1}$ and $\varphi=\pi$. In panels (c) and (d) the vertical dashed lines indicate the critical temperature of S$_1$. All the results have been obtained by setting the normal-state resistance of the JJ $R_{\rm j}=1$ k$\Omega$.}
\label{Fig1}
\end{figure} 

\section{NDTC in a temperature-biased JJ}\label{sect2}

We shall start, first of all, by considering a JJ consisting of two different superconductors (S$_1$ and S$_2$) coupled by a thin insulating layer (I), as depicted in Fig.~\ref{Fig1}(a). If we set S$_1$ at the temperature $T_1$ and S$_2$ at $T_2$, with $T_1>T_2$, the electronic heat current flowing through the junction from S$_1$ to S$_2$ can be expressed as:\cite{MakiGriffin,Guttman,Zhao1,Zhao2,GiazottoAPL}
\begin{equation}
J_{\rm S_1S_2}(T_1,T_2,\varphi)=J_{\rm qp}(T_1,T_2)-J_{\rm int}(T_1,T_2)\rm{cos}\varphi . \label{JSIS}
\end{equation}
Here, the first term accounts for the heat carried by quasiparticles, $J_{\rm qp}(T_1,T_2)=(2/e^{2}R_{\rm j}) \displaystyle\int_{0}^{\infty}  \! \epsilon \mathcal{N}_1(\epsilon , T_1)\mathcal{N}_2(\epsilon , T_2) \allowbreak [f(\epsilon ,T_1)-f(\epsilon , T_2)] \, \mathrm{d}\epsilon$, where $\mathcal{N}_{1,2}(\epsilon , T_{1,2})=|\Re[(\epsilon+i\Gamma_{1,2})/\sqrt{(\epsilon+i\Gamma_{1,2})^2-\Delta_{1,2}^2(T_{1,2})}]|$ are the smeared (if $\Gamma_{1,2} \neq 0$) normalized Bardeeen-Cooper-Schrieffer (BCS) densities of states (DOSs) of the superconductors,\cite{Dynes} $f(\epsilon ,T_{1,2})=[1+\text{exp}(\epsilon/k_{\rm B} T_{1,2})]^{-1}$ is the Fermi-Dirac distribution, $\Delta_{1,2}(T_{1,2})$ are the temperature-dependent energy gaps,\cite{Tinkham} $R_{\rm j}$ is the tunnel junction normal-state resistance, $e$ is the electron charge and $k_{\rm B}$ is the Boltzmann constant. Unless specified otherwise, in the following we will set $\Gamma_{1,2}= \gamma_{1,2} \Delta_{1,2}(0)$, with $\gamma_1=\gamma_2=\gamma=10^{-4}$, which describes realistic superconducting tunnel junctions.\cite{MartinezNatRect,PekolaPRL2004,PekolaPRL2010} Furthermore, we shall assume, for clarity, that $\delta = \Delta_2(0)/\Delta_1(0) \leq 1$.

The second component of Eq.~\eqref{JSIS} stands for the phase-coherent part of the heat current, which originates from energy-carrying tunneling processes involving concomitant destruction and creation of Cooper pairs on different sides of the junction.\cite{MakiGriffin,Guttman} It is therefore regulated by the phase difference $\varphi$ between the superconducting condensates and it can be written as $J_{\rm int}(T_{1},T_2)=(2/e^{2}R_{\rm j}) \displaystyle\int_{0}^{\infty}  \! \epsilon \mathcal{M}_1(\epsilon , T_1)\mathcal{M}_2(\epsilon , T_2) \allowbreak [f(\epsilon ,T_1)-f(\epsilon , T_2)] \, \mathrm{d}\epsilon$ (Ref. \citenum{GiazottoAPL}), where $\mathcal{M}_{1,2}(\epsilon,T_{1,2}) =\allowbreak |\Im[-i \Delta_{1,2}(T_{1,2})/\sqrt{(\epsilon\allowbreak +i\Gamma_{1,2})^2-\Delta_{1,2}^2(T_{1,2})}]|$ is the Cooper pair BCS DOSs in the superconductors.\cite{Barone} $J_{\rm int}$ represent the thermal counterpart of the "quasiparticle-pair interference" contribution to the \textit{charge} current tunneling through a JJ.\cite{Josephson,Harris,Barone,PopNature} Depending on $\varphi$, it can flow in opposite direction with respect to that imposed by the thermal gradient, but the total heat current $J_{\rm S_1S_2}$ still flows from the hot to the cold reservoir, thus preserving the second principle of thermodynamics. This was experimentally demonstrated in Ref. \citenum{GiazottoNature}.

Figure~\ref{Fig1}(b) shows the behavior of $J_{\rm S_1S_2}$ vs. $T_1$ for $T_2 = 0.01T_{\rm c1}$ ($T_{\rm c1}$ being the critical temperature of S$_1$) and $\delta = 0.75$. It appears evident how the variation of $\varphi$ can strongly influence the thermal transport through the JJ. First, let us focus on the case in which $\varphi=\pi/2$. In this condition, $J_{\rm S_1S_2}$ becomes equal to $J_{\rm qp}$, which presents a sharp peak at $T_1 \simeq 0.77 T_{\rm c1}$, due to the matching of singularities in the superconducting DOSs $\mathcal{N}$ when $\Delta_1(T_1) = \Delta_2(T_2)$. At higher values of $T_1$, $\Delta_1(T_1) <\Delta_2(T_2)$ and the energy transmission through the junction is reduced, thus originating an effect of NDTC. This feature is the analogue of the well-known singularity-matching peak (SMP) usually observed in the quasiparticle current flowing through a voltage-biased S$_1$IS$_2$ junction.\cite{Barone} Yet, in the thermal configuration, the effect of NDTC can be enhanced or reduced by the presence of $J_{\rm int}$ as determined by the value of $\varphi$
. At $\varphi=0$ the SMP is perfectly canceled by the coherent component of the heat current, while at $\varphi=\pi$ it becomes almost doubled and an additional NDTC feature appears, owing to the gradual suppression of $J_{\rm int}$ as $T_1$ approaches $T_{\rm c1}$. This results in a remarkable peak-to-valley ratio of $\simeq 3.1$. The behaviour of $J_{\rm int}$ is due to the singularity of $\mathcal{M}$ at $\epsilon=\Delta$ that perfectly corresponds to the one in $\mathcal{N}$, creating a sort of resonance between quasiparticle and pair tunneling.\cite{Harris,Barone}

The effect of NDTC depends also on the amplitude of $\Delta_2(T_2)$, as shown in the contour plot of Fig.~\ref{Fig1}(c). As $T_2$ is increased, the position of the SMP moves towards higher values of $T_1$ and its amplitude gradually decreases. It is worth noting that while the NDTC effect extends from the SMP to $T_{\rm c1}$ if we vary $T_1$ and keep $T_2$ fixed, it is much more localized in the proximity of the SMP if we vary $T_2$ and keep $T_1$ fixed. This will be important to understand the performances of different configurations for a superconducting thermal transistor (see Sect.~\ref{sect8}).

Finally, Fig.~\ref{Fig1}(d) displays the impact of $\delta$ on the region of NDTC. As $\Delta_2(0)$ becomes more similar to $\Delta_1(0)$ the extension of the NDTC region increases to the detriment of its amplitude. Therefore, the best configuration results to be the one with $\delta \simeq 0.75$.
 
\section{Behavior of the Josephson current}\label{sect3}

\begin{figure}[t]
\centering
\includegraphics[width=\columnwidth]{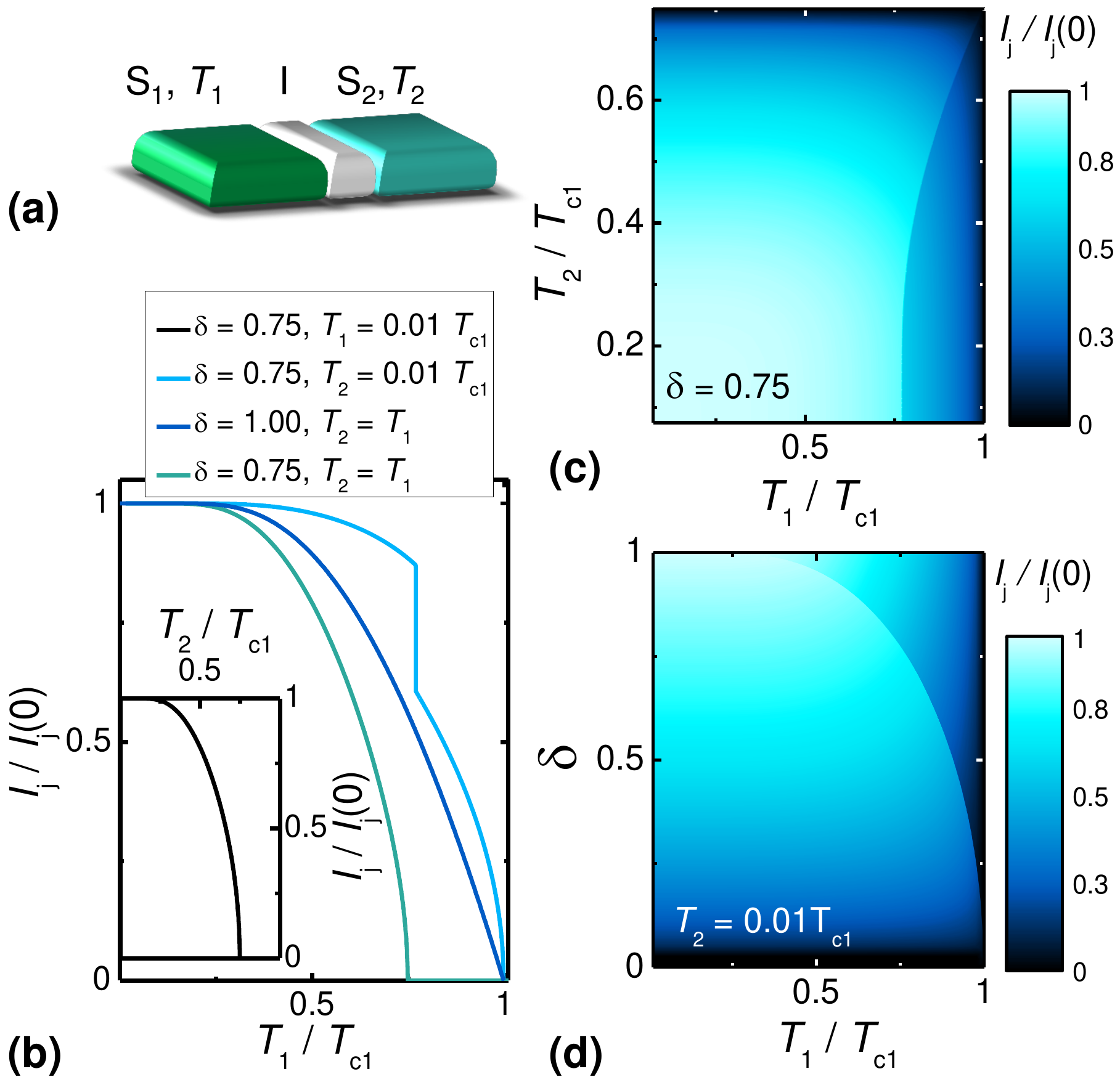}
\caption{Non-dissipative charge transport through a JJ. (a) Schematic configuration of the same JJ depicted in Fig.~\ref{Fig1}(a). (b) Normalized Josephson critical current $I_{\rm j}$ vs. $T_1$ for different configurations of $T_2$ and $\delta$. Inset: $I_{\rm j}$ as a function of $T_2$ for $\delta=0.75$ and $T_1=0.01T_{\rm c1}$. (c) Contour plot showing $I_{\rm j}$ vs. $T_1$ and $T_2$ for $\delta=0.75$. (d) Contour plot showing $I_{\rm j}$ vs. $T_1$ and $\delta$ for $T_2=0.01T_{\rm c1}$.
}
\label{Fig2}
\end{figure}

Before proceeding in the analysis of the possible ways to exploit NDTC, we first consider the electrical behavior of the JJ. The latter, as we shall argue, presents interesting features and can be used to probe the electronic temperature in a superconductor.

The system described in the previous section can support a non-dissipative Josephson current that follows the well-known expression:\cite{Josephson}
\begin{equation}
I_{\rm 0}(T_1,T_2)=I_{\rm j}(T_1,T_2)\rm{sin}\varphi,
\end{equation}
where $I_{\rm j}$ is the critical current of the JJ, which can be evaluated with the generalized Ambegaokar-Baratoff relation:\cite{GiazottoJap,Tirelli}
\begin{align}
I_{\rm j}(T_1,T_2)=&\frac{1}{2eR_{\rm j}}\left|\displaystyle\int_{-\infty}^{\infty} \mathrm{d}\epsilon\lbrace \textbf{f}(\epsilon,T_1)\Re[\mathcal{F}_1(\epsilon,T_1)]\Im[\mathcal{F}_2(\epsilon,T_2)]\right.\notag\\ 
&\left.+\textbf{f}(\epsilon,T_2)\Re[\mathcal{F}_2(\epsilon,T_2)]\Im[\mathcal{F}_1(\epsilon,T_1)]\vphantom{\frac{1}{2}}\right|.
\end{align}
Here, $\textbf{f}(\epsilon,T_{1,2})=\mathrm{tanh}(\epsilon/2k_{\rm B}T_{1,2})$ and $\mathcal{F}_{1,2}(\epsilon,T_{1,2})=\Delta_{1,2}/\sqrt{(\epsilon+i\Gamma_{1,2})^2-\Delta_{1,2}^2(T_{1,2})}$ are the anomalous Green's functions in the superconductors.\cite{Barone}

Figure~\ref{Fig2}(b) displays $I_{j}$ as a function of $T_1$ for three representative configurations of the JJ. First we consider the case in which no temperature gradient is set across the junction: if $\delta=1$, we recover the conventional result by Ambegaokar-Baratoff,\cite{AB} i.e. $I_{\rm j}=(\pi \Delta/2eR_{\rm j})\mathrm{tanh}(\Delta/2k_{\rm B}T)$ vanishing at $T_{\rm c1}$ with a finite slope. On the other hand, if $\delta<1$ the critical current goes to zero at $T_{\rm c2}$ with an infinite slope, following the BCS temperature-dependence of $\Delta_2$. More interestingly, if we fix $T_2$ and we let only $T_1$ vary, we obtain a sharp jump of $I_{\rm j}$ at $T_1\simeq 0.77T_{\rm c1}$ for $\delta=0.75$. This feature stems again from the alignment of the singularities in the Green's functions $\mathcal{F}$ at $\epsilon=\Delta$ when $\Delta_1(T_1)=\Delta_2(T_2)$, and to our knowledge it has never been observed so far. As shown in the inset of Fig.~\ref{Fig2}(b), if we vary $T_2$ and keep $T_1=0.01T_{\rm c1}$ the critical current decreases monotonically and without jumps, since in this configuration the condition $\Delta_1(T_1)=\Delta_2(T_2)$ is never met. The occurrence of this condition is mapped in the contour plots of Figs.~\ref{Fig2}(c) and~\ref{Fig2}(d), which are the equivalent of those shown previously in Figs.~\ref{Fig1}(c) and~\ref{Fig1}(d).

The above analysis confirms that a JJ can easily serve as a non-dissipative thermometer for the electronic temperature of a superconducting electrode above $\simeq 0.4 T_{\rm c}$.\cite{GiazottoRev} Since the NDTC effect occurs at temperatures relatively close to the critical one, for our purposes this kind of thermometry would represent a good alternative to more conventional methods, which are focused on the quasiparticle transport.\cite{GiazottoRev,Timofeev}

\section{Phase-bias of the JJ}\label{sect4}

\begin{figure}[t]
\centering
\includegraphics[width=\columnwidth]{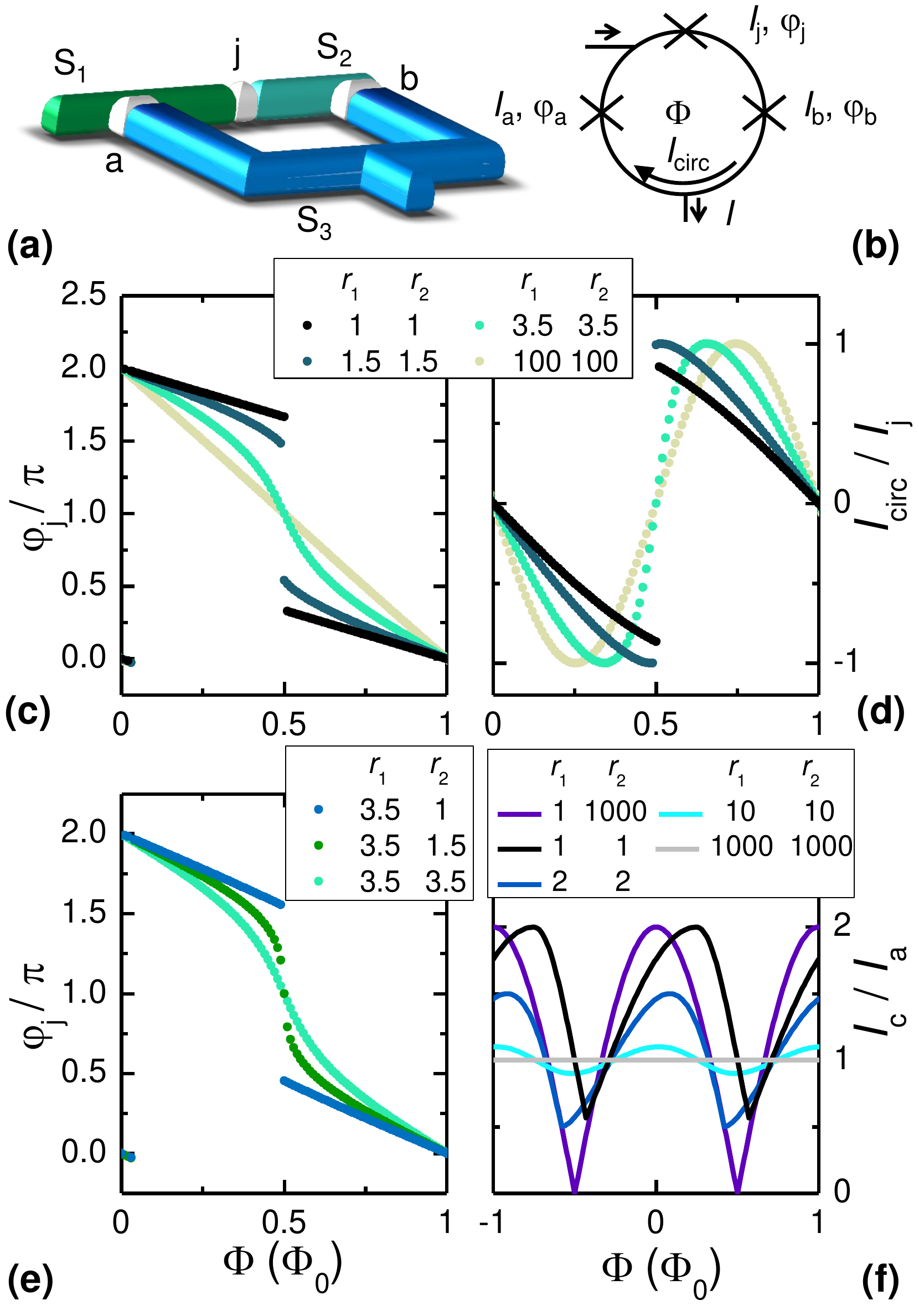}
\caption{Phase-biasing a JJ by means of a three-junction SQUID. (a) Pictorial representation of the interferometer. (b) Circuital scheme of the device. (c) Magnetic-flux dependence of the phase difference $\varphi_{\rm j}$ across the junction j for symmetric values of $r_1=I_{\rm a}/I_{\rm j}$ and $r_2=I_{\rm b}/I_{\rm j}$. (d) Normalized circulating current $I_{\rm circ}$ vs. $\Phi$ for the same values of $r_1$ and $r_2$ displayed in panel (c). (e) Phase polarization of $\varphi_{\rm j}$ vs. $\Phi$ for asymmetric values of $r_1$ and $r_2$. (f) Magnetic-flux dependence of the normalized SQUID critical current $I_{\rm c}$ for several combinations of $r_1$ and $r_2$.}
\label{Fig3}
\end{figure}

In order to maximize the effect of NDTC, the JJ between S$_1$ and S$_2$ (that we will label as j in this section) must be biased at $\varphi=\pi$, as shown in Sect.~\ref{sect2}. Phase biasing of a JJ can be achieved, in general, through supercurrent injection or by applying an external magnetic flux.\cite{MartinezAPLrect} In our case, the ideal way to obtain a full control over $\varphi$ is to realize a "fake" radio frequency superconducting quantum interference device (rf SQUID), as depicted in Fig.~\ref{Fig3}(a). The superconducting electrodes S$_1$ and S$_2$ are connected to a third superconductor S$_3$ (possibly with $\Delta_3(0)>\Delta_1(0),\Delta_2(0)$ so to suppress heat losses) by means of two parallel JJs named a and b. The three superconductors form a loop with three JJs, two of which are in series on the same branch [see Fig.~\ref{Fig3}(b)]. As we shall argue, in order to obtain a $\pi$ polarization between S$_1$ and S$_2$, the junction j must be characterized by the lowest Josephson critical current in the SQUID, so that most of the phase drop occurs across this junction. 

The described interferometer is characterized by the following set of equations:
\begin{align}
\label{fluxquant}
&\varphi_{\rm a}-(\varphi_{\rm b}+\varphi_{\rm j})+2\pi \frac{\Phi}{\Phi_0}=2n\pi,\\
\label{kirchhoff}
&I=I_{\rm a}\mathrm{sin}\varphi_{\rm a}+I_{\rm b}\mathrm{sin}\varphi_{\rm b},\\
\label{currentconservation}
&I_{\rm j}\mathrm{sin}\varphi_{\rm j}=I_{\rm b}\mathrm{sin}\varphi_{\rm b},\\
\label{circulatingcurrent}
&I_{\rm circ}=\frac{1}{2}(I_{\rm a}\mathrm{sin}\varphi_{\rm a}-I_{\rm b}\mathrm{sin}\varphi_{\rm b}),
\end{align}
where $I_{\rm k}$ and $\varphi_{\rm k}$ are the Josephson critical current and phase difference for the k-th junction, with k$\,=\,$a,b,j, $\Phi$ is the external magnetic flux threading the loop, $\Phi_0\simeq 2 \times 10^{-15}$ is the superconducting flux quantum and $n$ is an integer. Equation~\eqref{fluxquant} establishes the flux-phase quantization along the loop, Eq.~\eqref{kirchhoff} expresses the Kirchhoff law for the total supercurrent $I$ flowing through the SQUID, Eq.~\eqref{currentconservation} imposes the current conservation in one branch of the interferometer and, finally, Eq.~\eqref{circulatingcurrent} describes the circulating supercurrent $I_{\rm circ}$.

As we shall explain in the following section, we can phase-bias the thermal transport through junction j by just applying an external magnetic flux piercing the loop of the SQUID. In this configuration, only a circulating supercurrent can flow along the loop and $I=0$. From Eqs.~\eqref{fluxquant} and~\eqref{currentconservation}, we can extract the following expressions for $\varphi_{\rm a}$ and $\varphi_{\rm b}$:
\begin{align}
\label{phia}
&\varphi_{\rm a}=(\varphi_{\rm b}+\varphi_{\rm j})+2\pi \frac{\Phi}{\Phi_0},\\
\label{phib}
&\varphi_{\rm b}=(-1)^m\mathrm{arcsin}\left(\frac{1}{r_2}\mathrm{sin}\varphi_{\rm j}\right)+m\pi,
\end{align}
where $r_2=I_{\rm b}/I_{\rm j}$ and $m=0,1$. If we substitute Eqs.~\eqref{phia} and~\eqref{phib} into Eq.~\eqref{kirchhoff}, we obtain two branches of solutions for $\varphi_{\rm j}$ depending on $m$. The correct physical values are those which minimize the Josephson free energy of the system $E_{\rm J}=E_{\rm J}^{\rm a}+E_{\rm J}^{\rm b}+E_{\rm J}^{\rm j}$, with $E_{\rm J}^{\rm k}=(\Phi_0 I_{\rm k}/2\pi)(1-\mathrm{cos}\varphi_{\rm k})$.\cite{Tinkham} 

Figures~\ref{Fig3}(c) and~\ref{Fig3}(d) show the magnetic-flux dependence of $\varphi_{\rm j}$ and $I_{\rm circ}$ for different values of $r_1$ and $r_2$, where $r_1=I_{\rm a}/I_{\rm j}$. As $r_1$ and $r_2$ increase (i.e. as $I_{\rm j}$ becomes smaller than $I_{\rm a}$, $I_{\rm b}$), $\varphi_{\rm j}$ is able to reach the values around $\pi$ more smoothly and the $I_{\rm circ}$ characteristic becomes more sinusoidal, like in a standard rf SQUID. The obtained results reveal that the threshold to obtain a continuous $\pi$ polarization (without abrupt switches) is $r_1=r_2\geq 2.5$. Moreover, if we introduce an asymmetry between $I_{\rm a}$ and $I_{\rm b}$ above 40\%, the jump in the $\varphi_{\rm j}$ polarization curve reappears, as shown in Fig.~\ref{Fig3}(e). It is also worth noting that when $\varphi_{\rm j}=\pi$, we have $\varphi_{\rm a}=\varphi_{\rm b}=0$.

To conclude this section, we discuss the magnetic interference pattern of the SQUID total critical current $I_{\rm c}$, which represents the simplest measurement to characterize the interferometer. To obtain $I_{\rm c}(\Phi)$ we substitute again Eqs.~\eqref{phia} and~\eqref{phib} into Eq.~\eqref{kirchhoff} and we maximize the value of $I$ with respect to $\varphi_{\rm j}$. As previously mentioned, the correct solution is the one corresponding to the minimum of the Josephson energy. The resulting behavior of $I_{\rm c}$ vs. $\Phi$ is shown in Fig.~\ref{Fig3}(d), where we recognize three limit cases: first, if $(r_1,r_2)=(1,1000)$, i.e. $I_{\rm b}\gg I_{\rm a,j}$, the junction b becomes almost completely transparent, leaving just the junctions a and j to define a symmetric direct-current SQUID with the conventional pattern $\propto \vert \mathrm{cos}(\pi \Phi/\Phi_0)\vert$. On the other hand, if $(r_1,r_2)=(1,1)$, that is $I_{\rm a}=I_{\rm b}=I_{\rm j}$, the three-junction SQUID is completely symmetric and $I_{\rm c}$ presents a skewed pattern that never vanishes. Lastly, when $(r_1,r_2)=(1000,1000)$, i.e. $I_{\rm a}=I_{\rm b}\gg I_{\rm j}$, the junctions a and b become almost transparent with respect to junction j, thus forming a true rf SQUID. The latter is characterized by an almost constant $I_{\rm c}$, since the branch with only the transparent junction a shunts the circuit. We also notice that for $(r_1,r_2)\geq(2.5,2.5)$ the $I_{\rm c}$ characteristic loses the cusped minima and progressively turns into a sinusoid with reduced contrast. 

\section{NDTC and thermal memory in a heat tunnel diode}\label{sect5}

\begin{figure}[t]
\centering
\includegraphics[width=\columnwidth]{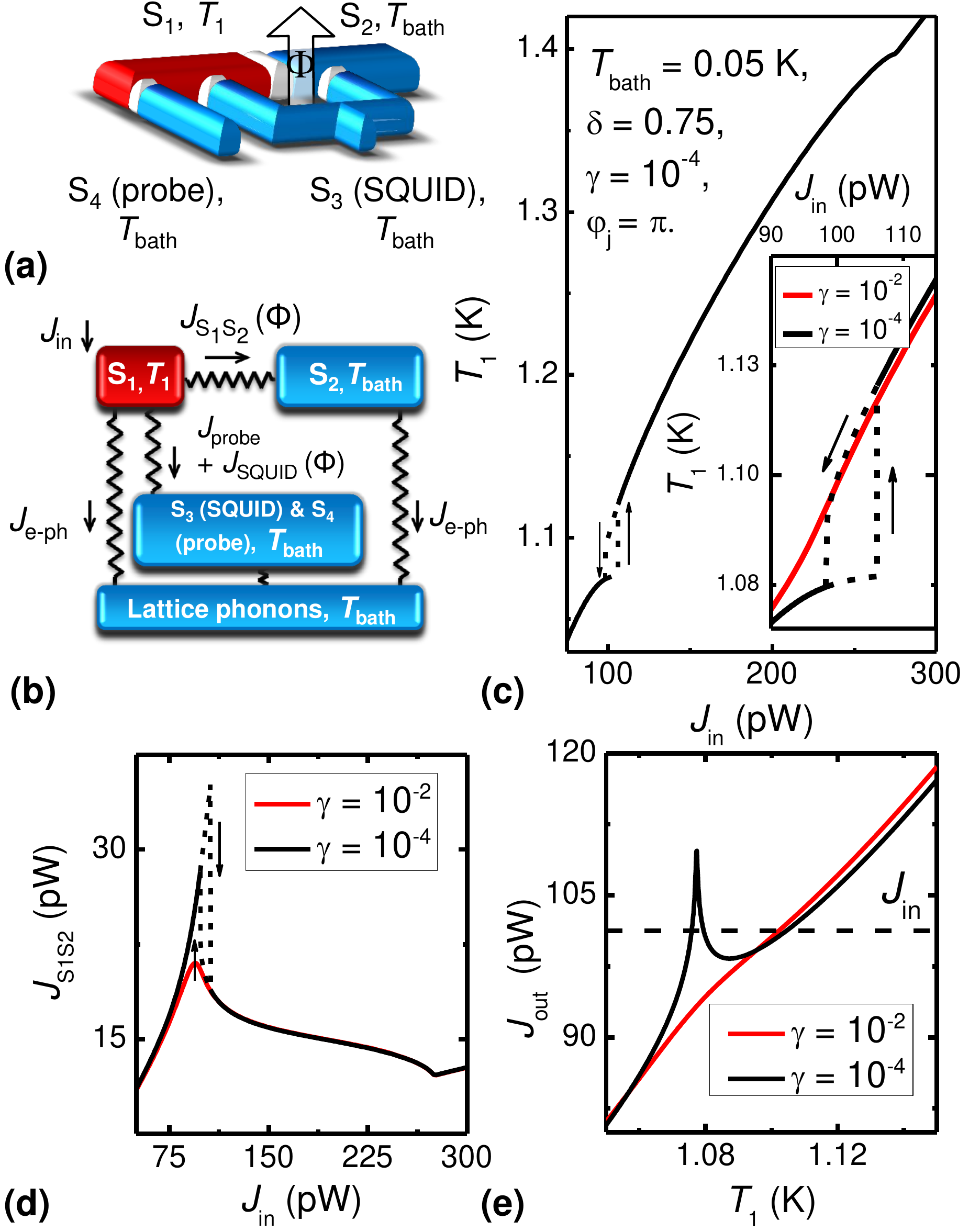}
\caption{Performance of the tunnel heat diode (design 1). (a) Pictorial representation of the device. (b) Thermal model outlining the relevant heat exchange mechanisms present in our tunnel diode. Arrows indicate heat current directions for the operating device, that is, when $T_1$ > $ T_{\rm bath}$ (see text).  (c) Calculated electronic temperature $T_1$ vs. the injected power $J_{\rm in}$ for $\gamma=10^{-4}$.  Inset: zoom of the hysteretic region of $T_1$ for two values of $\gamma$. The arrows near the curves indicate the $J_{\rm in}$ sweep direction. (d) Electronic heat current $J_{\rm S_1S_2}$ vs. $J_{\rm in}$ for two values of $\gamma$. (e) Input and output heat currents $J_{\rm in}$ (dashed line) and $J_{\rm out}$ (solid line) vs. $T_1$ for the same values of $\gamma$ shown in the other panels. All the results have been obtained at a bath temperature $T_{\rm bath}=50$ mK and for $\delta=0.75$, $\varphi_{\rm j}=\pi$ and $\varphi_{\rm a}=\varphi_{\rm b}=0$. We assumed that S$_1$, S$_3$ and S$_4$ are composed of aluminum with $T_{\rm c}=1.4$ K. The normal-state resistances are $R_{\rm j}=2$ k$\Omega$ and $R_{\rm a}=R_{\rm probe}=500\;\Omega$, whereas the volume of S$_1$ is $\mathcal{V}_1=1\times 10^{-19}$ m$^{3}$. For completeness, in the thermal model we also included two superconducting probes tunnel-coupled to S$_1$ acting as Josephson thermometers (see Sect.\ref{sect3}) with a normal-state resistance $R_{\rm thermo}=2$ k$\Omega$ for each junction (not shown).}
\label{Fig4}
\end{figure}

We now have all the elements necessary to envision a realistic caloritronic device able to provide a measurable NDTC, i.e. the thermal analogue of the electric tunnel diode. As any direct measurement of the heat current is unfeasible, the design of the thermal tunnel diode must be conceived to manifest sizable effects in the temperature of the electrodes. The simplest geometry for such a device consists in connecting a superconducting lead S$_3$ to S$_1$ and S$_2$, forming the three-junction SQUID described in the previous section [see Fig.~\ref{Fig4}(a)]. Furthermore, a superconducting probe S$_4$ tunnel-coupled to S$_1$ would offer the possibility to investigate the electrical transport through the device (see Sects.~\ref{sect3} and~\ref{sect4}). We emphasize that our analysis is focused on the heat carried by electrons only. We assume that lattice phonons present in every part of our structure are fully thermalized with the substrate phonons residing at the bath temperature $T_{\rm bath}$, thanks to the vanishing Kapitza resistance between thin metallic films and the substrate at low temperatures.\cite{GiazottoNature,MartinezNature,MartinezNatRect,FornieriNature,Wellstood} If we inject a Joule power $J_{\rm in}$ into S$_1$, we can raise its electronic temperature $T_1$ significantly above $T_{\rm bath}$,\cite{Wellstood} generating a thermal gradient across the device. This hypothesis is expected to hold because the other electrodes of the tunnel diode (S$_2$, S$_3$ and S$_4$) can be designed to extend into large-volume leads, providing efficient thermalization of their quasiparticles at $T_{\rm bath}$. This thermal gradient originates a finite heat current $J_{\rm S_1S_2}$, which displays a remarkable NDTC effect, as shown in Sect.~\ref{sect2}.

In order to predict the behavior of the heat tunnel diode, we formulate a thermal model accounting for all the predominant heat exchange mechanisms present in the structure. The model is sketched in Fig.~\ref{Fig4}(b), where $J_{\rm SQUID}$ and $J_{\rm probe}$ are the electronic heat currents flowing from S$_1$ to S$_3$ and S$_4$, respectively, through two JJs characterized by normal-state resistances $R_{\rm a}$ and $R_{\rm probe}$. Furthermore, we take into account the energy relaxation due to the electron-phonon coupling $J_{\rm e-ph}$, which in a superconductor at temperature $T$ can be expressed as:\cite{Timofeev}
\begin{align}
J_{\rm{e-ph}}(T,T_{\rm bath})=&-\frac{\Sigma \mathcal{V}}{96 \zeta (5)k_{\rm B}^5}\int_{-\infty}^{\infty} \mathrm{d}E E \int_{-\infty}^{\infty} \mathrm{d}\rm{\epsilon}\rm{\epsilon}^2 \rm{sgn}(\rm{\epsilon})\notag \\
&\times L(E,E+\epsilon,T)\left\lbrace\mathrm{coth}\left(\frac{\mathrm{\epsilon}}{2k_{\rm B}T_{\rm bath}}\right)\right.\notag \\
&\times [\textbf{f}(E,T)-\textbf{f}(E+\epsilon,T)]\notag \\
&\left. -\textbf{f}(E,T) \textbf{f}(E+\epsilon,T)+1 \vphantom{\frac{1}{2}} \right\rbrace.
\end{align}
Here, $\Sigma$ is the material-dependent electron-phonon coupling constant, $\mathcal{V}$ is the volume of the superconducting electrode and $L(E,E',T)=\mathcal{N}(E,T)\mathcal{N}(E',T)[1-\Delta^2(T)/(EE')]$. Therefore, the steady-state electronic temperature $T_1$ can be calculated as a function of $J_{\rm in}$ by solving the following energy balance equation:
\begin{align}
J_{\rm in}=\,&J_{\rm out}\notag\\
=\,&J_{\rm S_1S_2}(T_1,T_{\rm bath})+J_{\rm e-ph}(T_1,T_{\rm bath})\notag\\
\label{energybalance}
&+J_{\rm SQUID}(T_1,T_{\rm bath})+J_{\rm probe}(T_1,T_{\rm bath}),
\end{align}
which imposes that the sum of all the incoming ($J_{\rm in}$) and outgoing ($J_{\rm out}$) heat currents for S$_1$ must be equal to zero. The resulting trend of $T_1$ vs. $J_{\rm in}$ is shown in Fig.~\ref{Fig4}(c), where we set $T_{\rm bath}=50$ mK, $\delta=0.75$, $\varphi_{\rm j}=\pi$ and $\varphi_{\rm a}=0$. The normal-state resistances were designated to be $R_{\rm j}=2$ k$\Omega$ and $R_{\rm a}=R_{\rm probe}=500\;\Omega$. We also assumed that S$_1$, the SQUID and the probe are composed of aluminum (Al) with $T_{\rm c}=1.4$ K and $\Sigma=3\times 10^8$ WK$^{-5}$m$^{-3}$ (Ref. \citenum{GiazottoRev}), whereas the volume of S$_1$ is $\mathcal{V}_1=1\times 10^{-19}$ m$^{3}$. In order to obtain a proper value of $\delta$, S$_2$ can be realized as a bilayer of a normal metal in clean contact with a superconductor: owing to the inverse proximity effect, $\Delta$ and $T_{\rm c}$ can be manipulated at will by varying the thicknesses of the layers.\cite{Martinis,Brammertz}

The calculated results present two prominent features: at $J_{\rm in}\simeq 100$ pW the slope of $T_1$ suddenly increases and a region of thermal hysteresis appears [see Fig.~\ref{Fig4}(c)]. As a matter of fact, the $T_1$ curve creates a loop instead of retracing its path for increasing and decreasing $J_{\rm in}$, showing bi-stable temperature states for a given input power. Both the features are indirect evidences of NDTC in $J_{\rm S_1S_2}$, which is displayed in Fig.~\ref{Fig4}(d). In particular, the increase in the derivative of $T_1$ corresponds to the onset of the NDTC regime, in which S$_1$ results to be more isolated from S$_2$ and gets heated more efficiently by the injection of $J_{\rm in}$. The end of the NDTC region coincides with the transition of S$_1$ into a normal metal at $1.4$ K, where $T_1$ shows a cusp. Even more interesting, for small values of $\gamma$ the SMP in $J_{\rm S_1S_2}$ generates an hysteresis in the $T_1$ curve, as highlighted in the inset of Fig.~\ref{Fig4}(c). This effect can be easily understood by plotting $J_{\rm in}$ and $J_{\rm out}$ vs. $T_1$, as displayed in Fig.~\ref{Fig4}(e) with a dashed and a solid line, respectively. In the graph, the intersections between $J_{\rm in}$ and $J_{\rm out}$ are indicating the possible solutions for $T_1$. When $\gamma=10^{-4}$, three solutions are visible for $98\,$pW$\,\lesssim J_{\rm in}\lesssim 106\,$pW, of which only two are in the positive slope parts of the $J_{\rm out}$ curve and are hence stable operating points of the device. 
On the contrary, if $\gamma$ is increased, the SMP in $J_{\rm S_1S_2}$ becomes broadened [see Fig.~\ref{Fig4}(d)], $J_{\rm out}$ turns into a monotonic function and Eq.~\eqref{energybalance} has therefore a single solution for $T_1$ in the whole range of $J_{\rm in}$.

The region of thermal hysteresis can be used to realize a thermal memory device, in analogy to what has been done in Ref. \citenum{XieAFM}. Indeed, in this region S$_1$ can reside at two different temperatures $T_{\rm high}$ and $T_{\rm low}$ for a given $J_{\rm in}$. These can be considered as the logical Boolean units 1 ($=T_{\rm high}$) and 0 ($=T_{\rm low}$) to store and read thermal information on the tunnel diode. In order to perform a cycle of writing and reading, we define $J_{\rm read}\simeq$ 102 pW in the middle of the hysteretic regime [see the inset of Fig.~\ref{Fig4}(c)], whereas we label $J_{\rm high}=110$ pW and $J_{\rm low}=95$ pW outside the boundaries of the multi-valued region. In this way, we can write 1 or 0 by setting $J_{\rm in}=J_{\rm high}$ or $J_{\rm in}=J_{\rm low}$, respectively, and afterward read the stored information by applying $J_{\rm in}=J_{\rm read}$. The performance and repeatability of this process is certainly improved if the temperature difference $\delta T=[T_{\rm high}-T_{\rm low}]_{J_{\rm in,read}}$ is maximized, in order to reduce the number of errors caused by noise and fluctuations. For the chosen $R_{\rm j}$, $\delta T=32$ mK, but this value can be increased up to $56$ mK by reducing $R_{\rm j}$ down to 1.1 k$\Omega$ (the latter being the minimum normal-state resistance that preserves the $\pi$-polarization of $\varphi_{\rm j}$).

Finally, we spend a few words about the heat current noise that might affect the proposed system, leading to a reduced visibility of the hysteretic regime. We assume to inject $J_{\rm in}$ by means of two superconducting probes S$_5$ tunnel-coupled to S$_1$ in order to form a S$_5$IS$_1$IS$_5$ junction.\cite{Timofeev} If we apply a voltage $V>2(\Delta_1+\Delta_5)/e$, we dissipate a Joule power in S$_1$, which represents the main source of noise in our system. The noise spectral density associated to $J_{\rm in}$ is detailed in Ref. \citenum{GolubevPRB} and can be estimated of the order of $10^{-17}\div 10^{-16}$ W/Hz$^{1/2}$. In the present setup, the admitted frequency band can extend up to a few MHz,\cite{PekolaPRL} leading to fluctuations amplitudes of $\sim 10^{-13}$ W, i.e. at least one order of magnitudes less than the power scale needed to control the hysteresis of our thermal tunnel diode. This estimation is confirmed experimentally by Ref. \citenum{FornieriNature}, where the visibility of features with an amplitude of a few mK in the interference pattern generated by a Josephson heat modulator corresponds to a sensitivity of $10^{-14}\div 10^{-13}$ W in terms of electronic heat currents.

\section{Alternative design of the heat tunnel diode}\label{sect6}

\begin{figure}[t]
\centering
\includegraphics[width=\columnwidth]{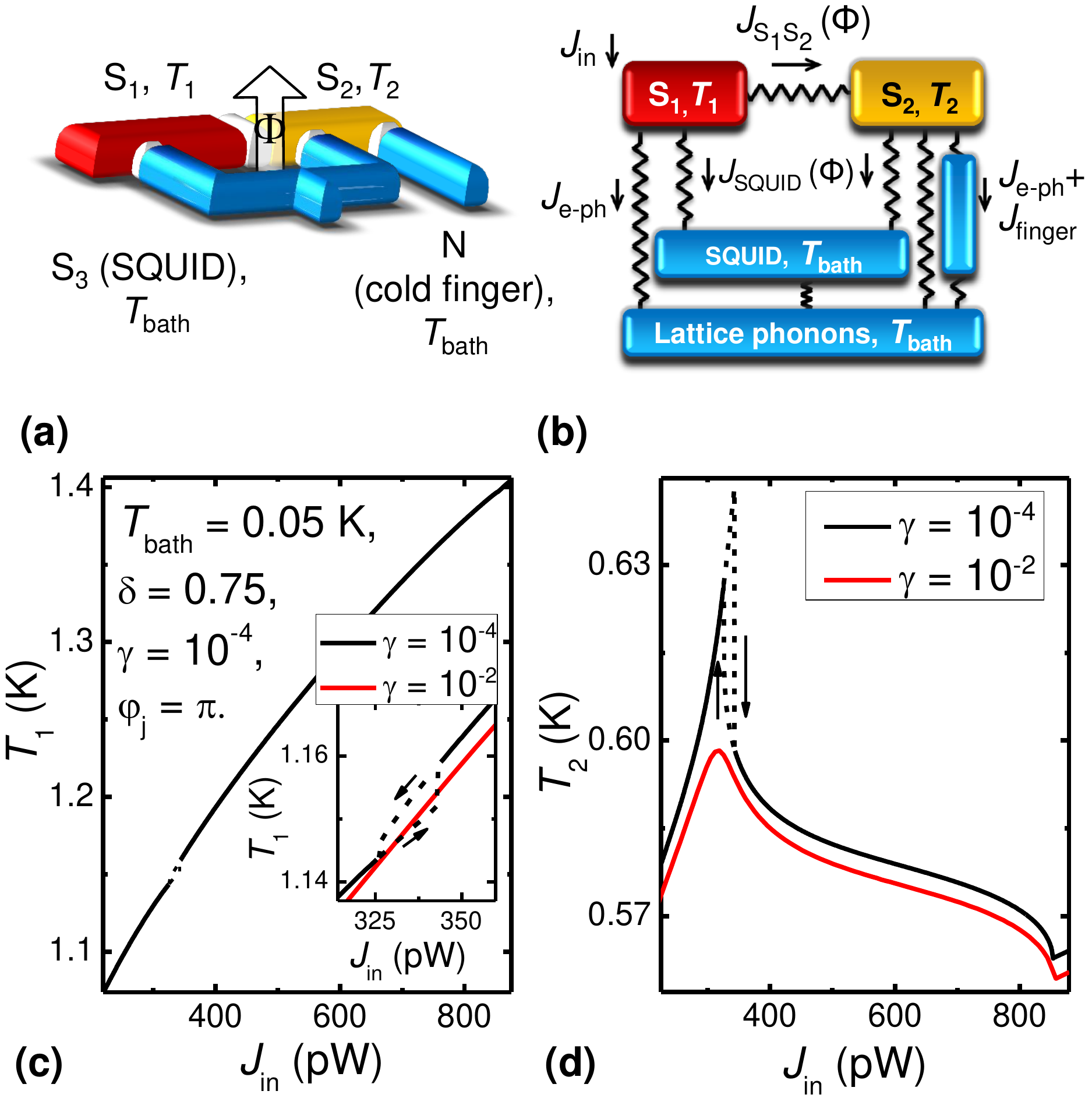}
\caption{Performance of the tunnel heat diode (design 2). (a) Pictorial representation of the device. (b) Thermal model outlining the relevant heat exchange mechanisms present in our tunnel diode when $T_1$ > $T_2$ > $ T_{\rm bath}$.  (c) Calculated electronic temperature $T_1$ vs. the injected power $J_{\rm in}$ for $\gamma=10^{-4}$.  Inset: zoom of the hysteretic region of $T_1$ for two values of $\gamma$. The arrows near the curves indicate the $J_{\rm in}$ sweep direction. (d) Electronic temperature $T_2$ vs. $J_{\rm in}$ for the same values of $\gamma$ shown in the inset of panel (c). All the results have been obtained at a bath temperature $T_{\rm bath}=50$ mK and for $\delta=0.75$, $\varphi_{\rm j}=\pi$ and $\varphi_{\rm a}=\varphi_{\rm b}=0$. We assumed that S$_1$ and S$_3$ are made of Al with $T_{\rm c}=1.4$ K. The normal-state resistances are $R_{\rm j}=2$ k$\Omega$ and $R_{\rm a}=R_{\rm b}=R_{\rm finger}=500\;\Omega$, whereas the volume of S$_1$ and S$_2$ are $\mathcal{V}_1=5\times 10^{-20}$ m$^{3}$ and $\mathcal{V}_2=1\times 10^{-19}$ m$^{3}$, respectively. For completeness, in the thermal model we also included two pairs of superconducting probes tunnel-coupled to S$_1$ and S$_2$ acting as Josephson thermometers (see Sect.\ref{sect3}) with a normal-state resistance $R_{\rm thermo}=2$ k$\Omega$ for each junction (not shown).}
\label{Fig5}
\end{figure}

In this section, we briefly describe an alternative design of the tunnel heat diode, in which, as we shall show, the output temperature trend directly reflects the behavior of $J_{\rm S_1S_2}$. The basic elements of the device are displayed in Fig.~\ref{Fig5}(a), where we can notice two main differences with respect to the previous version of the tunnel diode. Firstly, S$_2$ is not completely thermalized at $T_{\rm bath}$, but its electronic temperature $T_2$ is floating and can be measured as a function of the Joule power $J_{\rm in}$ injected in S$_1$. Secondly, we added a normal-metal (N) electrode tunnel-coupled to S$_2$ acting as a cold finger, which permits to maintain a large temperature gradient between S$_1$ and S$_2$. The electronic heat current flowing through the S$_2$IN junction reads:\cite{GiazottoRev}
\begin{align}
J_{\rm finger}(T_2,T_{\rm bath})=&\frac{2}{e^{2}R_{\rm finger}} \displaystyle\int_{0}^{\infty}  \! \epsilon \mathcal{N}_2(\epsilon , T_2)\notag \\&[f(\epsilon ,T_2)-f(\epsilon , T_{\rm bath})] \, \mathrm{d}\epsilon,
\end{align}
where $R_{\rm finger}$ is the normal-state resistance of the junction.

The thermal model used to predict the performance of the device is shown in Fig.~\ref{Fig5}(b), from which we obtain two energy-balance equations describing the thermal response of the system vs. $J_{\rm in}$:
\begin{align}
\label{energybalance1}
J_{\rm in}=&J_{\rm S_1S_2}(T_1,T_2)+J_{\rm SQUID}(T_1,T_{\rm bath})\notag \\
&+J_{\rm e-ph}(T_1,T_{\rm bath}),\\
J_{\rm S_1S_2}(T_1,T_2)=&J_{\rm finger}(T_2,T_{\rm bath})+J_{\rm SQUID}(T_2,T_{\rm bath})\notag \\
&+J_{\rm e-ph}(T_2,T_{\rm bath}).
\end{align}
The resulting behaviors of $T_1$ and $T_2$ can be seen in Figs.~\ref{Fig5}(c) and~\ref{Fig5}(d). Here, we chose S$_1$ and S$_3$ made of Al and we set $T_{\rm bath}=50$ mK, $\delta=0.75$, $\varphi_{\rm j}=\pi$, $\varphi_{\rm a}=\varphi_{\rm b}=0$, $R_{\rm j}=2$ k$\Omega$ and $R_{\rm a}=R_{\rm b}=R_{\rm finger}=500\;\Omega$. We also assumed that S$_1$ and S$_2$ have volumes $\mathcal{V}_1=5\times 10^{-20}$ m$^3$ and $\mathcal{V}_2=1\times 10^{-19}$ m$^3$, respectively. The calculated $T_1$ vs $J_{\rm in}$ is almost identical to the curve shown in Fig.~\ref{Fig4}(c), even though all the features appear less evident owing to the reduced thermal gradient $T_1-T_2$, compared with the one obtained in the previous configuration. 
On the other hand, as $T_1$ increases, $T_2$ reaches a maximum and afterwards decreases until $T_1$ reaches $T_{\rm c1}$. We emphasize that this behavior represents the direct proof of the NDTC effect, which generates a reduction of $T_2$ that can be as large as 80 mK and a region of thermal hysteresis that depends on the value of $\gamma$.
It is therefore clear how this design, despite its slightly more complicated geometry and composition, might offer an indisputable direct evidence of NDTC and become an essential building block to realize a thermal transistor, as we shall argue in the next sections.

\section{Thermal switch and modulator}\label{sect7}

\begin{figure}[t]
\centering
\includegraphics[width=\columnwidth]{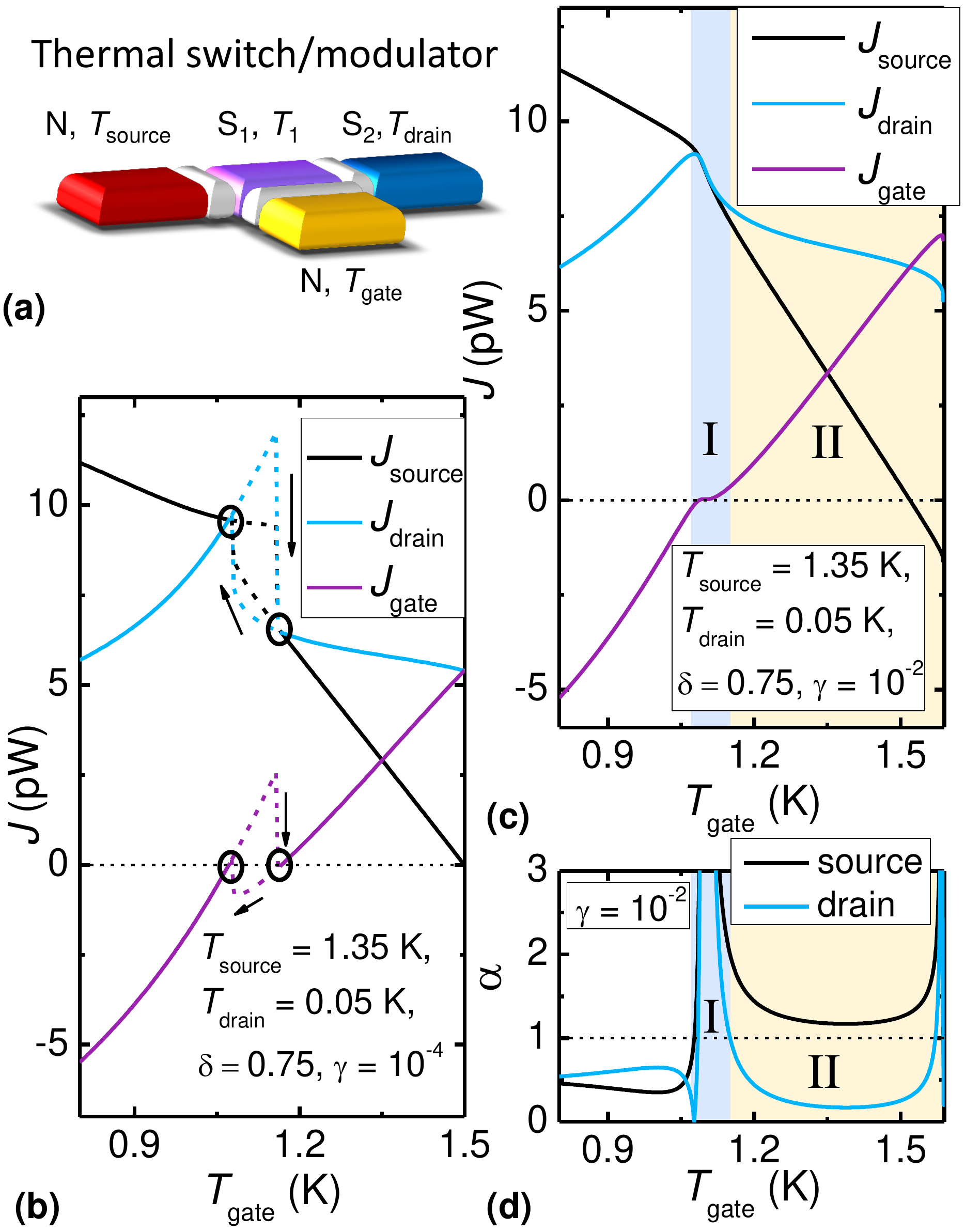}
\caption{Operation of a thermal switch and modulator. (a) Pictorial representation of the device. (b) Electronic heat currents $J_{\rm source}$, $J_{\rm drain}$ and $J_{\rm gate}$ vs. $T_{\rm gate}$ for $\gamma=10^{-4}$, $T_{\rm source}=1.35$ K, $T_{\rm drain}= 50$ mK, $\delta=0.75$, $R_{\rm source}=1$ k$\Omega$, $R_{\rm gate}=1$ k$\Omega$ and $R_{\rm drain}=5.25$ k$\Omega$. The arrows near the curves indicate the $T_{\rm gate}$ sweep direction, whereas the black circles indicate the stable working points of the thermal switch. (c) $J_{\rm source}$, $J_{\rm drain}$ and $J_{\rm gate}$ vs. $T_{\rm gate}$ for the same parameters used in panel (b), except for $\gamma=10^{-2}$ and $R_{\rm drain}=4.6$ k$\Omega$. In panels (b) and (c) the horizontal dashed lines outline $J=0$. (d) Amplification factors $\alpha_{\rm source,drain}$ vs. $T_{\rm gate}$ for the same parameters chosen in panel (c). The horizontal dashed line indicates $\alpha=1$. In panels (c) and (d), the shadowed regions I and II correspond to two different regimes of amplification (see text).}
\label{Fig6}
\end{figure}

Once we have the ability to control electronic heat currents with NDTC, it is natural to exploit it in order to realize a thermal transistor. Similarly to its electronic analogue, the thermal transistor consists of three terminals: the source, the drain and the gate, residing at temperatures $T_{\rm source}$, $T_{\rm drain}$ and $T_{\rm gate}$, respectively. The last terminal is a control knob that can tune the thermal flow across the device, offering also the opportunity to obtain \textit{heat amplification}. This is possible if the changes in the thermal current coming form the gate ($J_{\rm gate}$) can induce an even larger change in the currents flowing from the source to the drain.\cite{LiWangCasatiAPL,LiRev}

In order to envision the potential working operations of the transistor, we shall study two possible configurations of a three-terminal hybrid device. As shown in Fig.~\ref{Fig6}(a), in the first case two N electrodes play the roles of the source and the gate, while the central island and the drain consist in the S$_1$IS$_2$ junction that we analyzed in previous sections. In the following we will show that this structure can act as a thermal \textit{switch} and \textit{modulator}. On the other hand, if we connect a N gate to a S$_1$IS$_2$IN chain, we can obtain a thermal amplifier, as explained in the next section.

For simplicity, we set a fixed temperature gradient across the device, i.e. $T_{\rm source}>T_{\rm drain}$, and we assume that the device does not release energy to the environment. The latter hypothesis results to be accurate at low temperatures and for small volumes of the electrodes. Then, we analyze the behavior of the heat currents flowing out of the source and entering the drain ($J_{\rm source}$ and $J_{\rm drain}$, respectively) when we vary $T_{\rm gate}$. This is obtained by solving the following energy-balance equation:
\begin{align}
J_{\rm source}(T_{\rm source},T_1)+J_{\rm gate}(T_{\rm gate},T_1)=J_{\rm drain}(T_1,T_{\rm drain}),\label{Jconservation}
\end{align}
where $J_{\rm source}=(R_{\rm finger}/R_{\rm source})J_{\rm finger}$, $J_{\rm drain}=J_{\rm S_1S_2}$ and $J_{\rm gate}=(R_{\rm finger}/R_{\rm gate})J_{\rm finger}$, while $R_{\rm source}$, $R_{\rm gate}$ and $R_{\rm drain}$ are the normal-state resistances of the tunnel junctions connecting the central island to the other terminals of the device.

The results for $\gamma=10^{-4}$ are shown in Fig.~\ref{Fig6}(b), where we set $T_{\rm source}=1.35$ K, $T_{\rm drain}= 50$ mK, $\delta=0.75$, $R_{\rm source}=1$ k$\Omega$, $R_{\rm gate}=1$ k$\Omega$ and $R_{\rm drain}=5.25$ k$\Omega$. We notice that $J_{\rm gate}=0$ at $T_{\rm gate}=1.07,\;1.17$, pinpointing two stable working points where $J_{\rm source}=J_{\rm drain}$ [see the black circles in Fig.~\ref{Fig6}(b)]. These points can represent 1 and 0 Boolean states and do not depend on the history of the device. Therefore, our structure can work as a thermal switch.  

Additionally, this system can operate as a thermal modulator, as displayed in Fig.~\ref{Fig6}(c) for the same parameters listed above except for $R_{\rm drain}=4.6$ k$\Omega$ and $\gamma= 10^{-2}$. The latter value has been chosen to suppress the thermal hysteresis [see Sect.~\ref{sect5}] and simplify the following analysis on heat amplification. As a matter of fact, even at a first glance, it is possible to note that the device can remarkably reduce $J_{\rm source}$ and $J_{\rm drain}$ in a region where $J_{\rm gate}$ remains close to zero [see region I in Fig.~\ref{Fig6}(c)]. This behavior can be evaluated more quantitatively by defining the amplification factor:\cite{LiWangCasatiAPL}
\begin{equation}
\alpha_{\rm source,drain}=\left| \frac{\partial J_{\rm source,drain}}{\partial J_{\rm gate}}\right|=\left|\frac{g_{\rm source,drain}}{g_{\rm source}+g_{\rm drain}}\right|,\label{ampl}
\end{equation}
where we used Eq.~\eqref{Jconservation} and defined $g_{\rm source}=-\partial J_{\rm source}/\partial T_1$ and $g_{\rm drain}=\partial J_{\rm drain}/\partial T_1$ as the differential thermal conductances of the source and drain tunnel junctions. From Eq.~\eqref{ampl}, it is clear that $\alpha$ can be $>1$ only if one between $g_{\rm source}$ and $g_{\rm drain}$ is negative. In our case, the source is connected to the central island by means of the NIS$_1$ junction, which cannot show the NDTC effect and therefore we always have $g_{\rm source}>0$. Instead, the S$_1$IS$_2$ junction can generate $g_{\rm drain}<0$, as demonstrated in the previous sections.

The trend for $\alpha_{\rm source}$ and $\alpha_{\rm drain}$ vs. $T_{\rm gate}$ is shown in Fig.~\ref{Fig6}(d), where we can immediately distinguish two amplification regions I and II, shadowed in blue and yellow, respectively. In region I, i.e. for $1.08\,$K$\,\lesssim T_{\rm gate}\lesssim 1.15\,$K, the performance of the thermal modulator is ideal and both the amplification factors are $\gg 1$. As shown in Fig.~\ref{Fig6}(c), this corresponds to the regime characterized by:
\begin{equation}
\frac{\partial J_{\rm gate}}{\partial T_{\rm gate}}=\frac{\partial T_1}{\partial T_{\rm gate}}(g_{\rm source}+g_{\rm drain})\sim 0,\label{idealampl}
\end{equation}
where the identity between the first two sides has been obtained by using Eq.~\eqref{Jconservation}. Moreover, we have:
\begin{equation}
\frac{\partial T_1}{\partial T_{\rm gate}}=-g_{\rm source}\frac{\partial J_{\rm source}}{\partial T_{\rm gate}}>0,
\end{equation}
since $g_{\rm source}>0$ and $\partial J_{\rm source}/\partial T_{\rm gate}$ is negative in the whole range of operation [see Fig.~\ref{Fig6}(c)]. Thus, from Eq.\eqref{idealampl} we obtain that region I is characterized by $g_{\rm source}\sim -g_{\rm drain}$ and $\alpha_{\rm source,drain}\gg 1$. Yet, in the whole extension of region II, i.e. $1.15\,$K$\,\lesssim T_{\rm gate}\lesssim 1.59\,$K, the presence of the NDTC ($g_{\rm drain}<0$) still produces $\alpha_{\rm source}>1$, even though $J_{\rm gate}$ increases with a finite slope [see Fig.~\ref{Fig6}(c)]. 


This configuration would produce significant results also in a more realistic device that can release energy to the environment. As explained in the previous section, a N cold finger connected to S$_2$ would be able to maintain a relevant temperature gradient at the output of the transistor, leading to differences between the 1 and 0 states of the thermal switch exceeding 10 mK. Furthermore, it would be possible to obtain amplification factors $>1$ by limiting the impact of the electron-phonon coupling with small volumes of the electrodes, especially for what concerns S$_1$. 
\section{Thermal amplifier}\label{sect8}

\begin{figure}[t]
\centering
\includegraphics[width=\columnwidth]{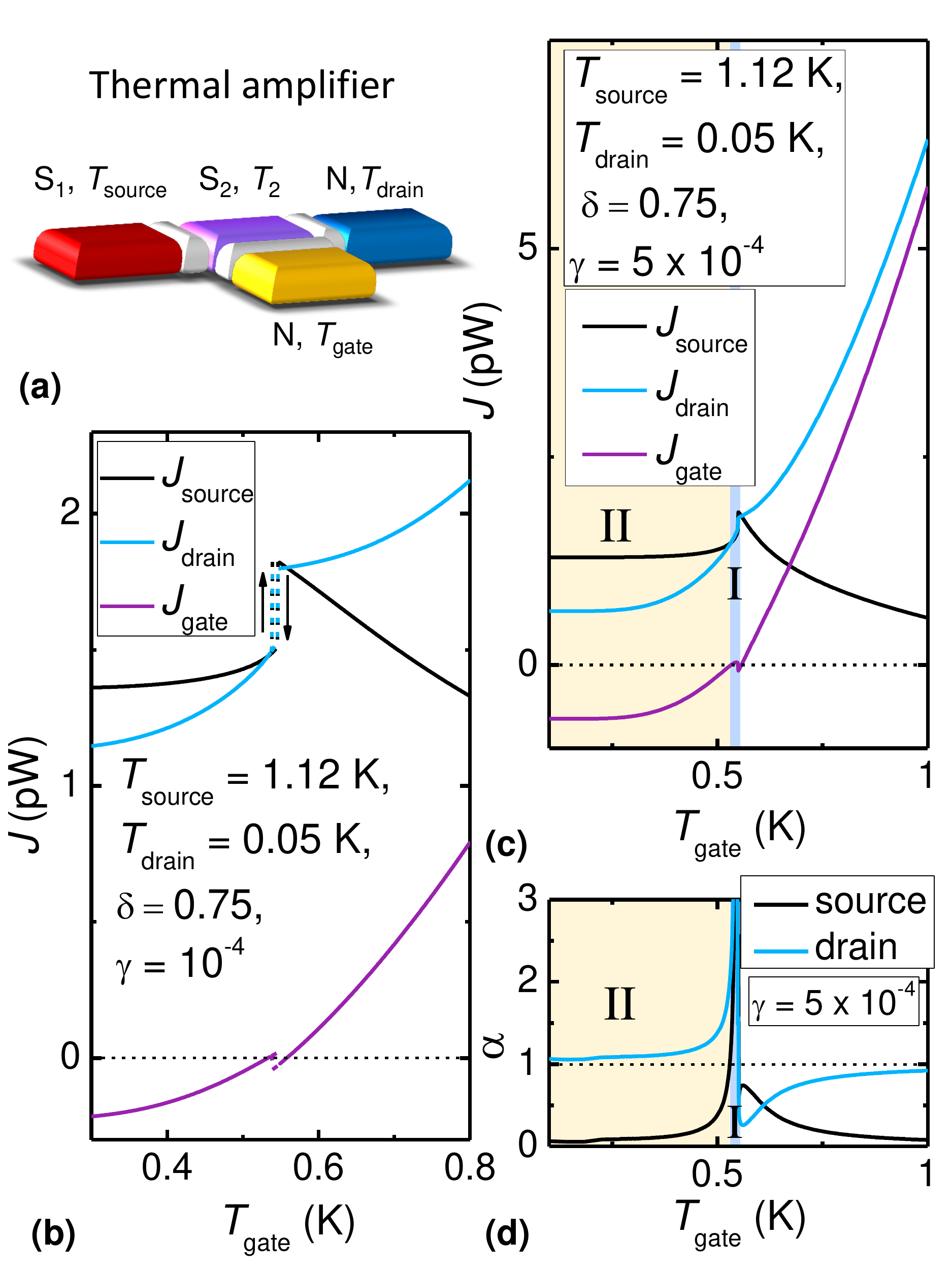}
\caption{Performance of a thermal amplifier. (a) Pictorial representation of the device. (b) Electronic heat currents $J_{\rm source}$, $J_{\rm drain}$ and $J_{\rm gate}$ vs. $T_{\rm gate}$ for $\gamma=10^{-4}$, $T_{\rm source}=1.12$ K, $T_{\rm drain}= 50$ mK, $\delta=0.75$, $R_{\rm source}=30$ k$\Omega$, $R_{\rm gate}=5$ k$\Omega$ and $R_{\rm drain}=1$ k$\Omega$. The arrows near the curves indicate the $T_{\rm gate}$ sweep direction. (c) $J_{\rm source}$, $J_{\rm drain}$ and $J_{\rm gate}$ vs. $T_{\rm gate}$ for the same parameters used in panel (b), except for $\gamma=5\times 10^{-4}$ and $R_{\rm gate}=1$ k$\Omega$. In panels (b) and (c) the horizontal dashed lines outline $J=0$. (d) Amplification factors $\alpha_{\rm source,drain}$ vs. $T_{\rm gate}$ for the same parameters chosen in panel (c). The horizontal dashed line indicates $\alpha=1$. In panels (c) and (d), the shadowed regions I and II correspond to two different regimes of amplification (see text).}
\label{Fig7}
\end{figure}

In this section, we briefly show the alternative configuration for a thermal transistor that can operate as a thermal \textit{amplifier}. The structure is shown in Fig.~\ref{Fig7}(a) and consists of two N electrodes acting as the gate and the drain, while S$_1$ and S$_2$ embody the source and the central island, respectively. If we take the same assumptions described in the previous section, we can calculate $J_{\rm source}$, $J_{\rm drain}$ and $J_{\rm gate}$ vs. $T_{\rm gate}$, as displayed in Fig.~\ref{Fig7}(b) (see the caption for the detailed list of parameters). 
It is easy to observe that the electronic heat current curves present a very small region of thermal hysteresis, which cannot be used to realize an effective thermal switch. As anticipated in Sect.~\ref{sect2}, the effect is indeed reduced with respect to that observed in previous devices, since in this case the NDTC is obtained by varying $T_2$ instead of $T_1$ and is consequently much more localized in the proximity of the SMP. Nevertheless, if we consider a higher value of $\gamma$ [see Fig.~\ref{Fig7}(c)], we can easily appreciate the amplification effect of our device. As we have noticed in the previous case, two different regimes I and II are visible [see Fig.~\ref{Fig7}(d)]: in region II, only $\alpha_{\rm drain}$ is raised above 1 by the NDTC, while in region I $(\partial J_{\rm gate}/\partial T_{\rm gate})\simeq 0$ and both $\alpha_{\rm source,drain}\gg 1$.

\section{Conclusions}\label{sect9}

In summary, we have analyzed the transport properties of a temperature-biased JJ composed by two different superconductors. From the point of view of charge transport, this S$_1$IS$_2$ junction can support a Josephson current, which present a remarkable jump when $\Delta_1(T_1)=\Delta_2(T_2)$. This feature has never been observed so far and would confirm the microscopic theory for a Josephson tunneling structure.\cite{AB,Harris,Barone} Moreover, the temperature-dependence of the Josephson current could provide a valid alternative to probe the electronic temperature in superconducting electrodes.

Foremost, on the thermal side, a temperature gradient imposed across the S$_1$IS$_2$ junction can produce a sizable effect of NDTC, which exhibits a maximum peak-to-valley ratio $\simeq 3.1$ in the transmitted electronic heat current when the phase difference between the superconducting condensates is $\pi$. This requirement can be fulfilled with the help of a three-junction SQUID controlled by an external magnetic flux. With these elements, we envisioned two different designs for a thermal tunnel diode, which could immediately be implemented to observe a temperature modulation as large as 80 mK due to the NDTC effect. Under proper conditions, this device would also produce a thermal hysteresis that might serve to store information in a solid-state memory device at cryogenic temperatures.

Finally, we showed the potential applications of NDTC into two versions of a thermal transistor. In the first case, the device can act as a thermal switch and modulator, while in the second configuration our three-terminal structure operates as a thermal amplifier. In both the schemes we are able to obtain a remarkable heat amplification in a wide range ($\sim 500$ mK) of the gate temperature. This result is a strict consequence of the NDTC, as predicted by Li \textit{et al}.\cite{LiWangCasatiAPL}

The proposed systems could be easily implemented by standard nanofabrication techniques and, combined with caloritronic interferometers\cite{GiazottoNature,MartinezNature,FornieriNature} and thermal diodes,\cite{MartinezNatRect} might represent the last missing pieces to complete the thermal reproduction of the most important electronic devices. Besides being relevant from a fundamental physics point of view, these structures would find immediate technological application as essential building blocks in solid-state thermal nanocircuits and in general-purpose cryogenic electronic applications requiring energy management.


We acknowledge the MIUR-FIRB2013–Project Coca (grant no. RBFR1379UX), the European Research Council under the European Union’s Seventh Framework Programme (FP7/2007-2013)/ERC grant agreement no. 615187 - COMANCHE and the European Union (FP7/2007-2013)/REA grant agreement no. 630925 – COHEAT for partial financial support.



\end{document}